\documentclass[aps,prl, reprint]{revtex4-1}
\usepackage{times}		
\usepackage{booktabs}
\usepackage{placeins}

\usepackage{amssymb,amstext,amsmath, epsf}
\usepackage{graphicx,graphics,epsfig,wrapfig}
\usepackage{listings}
\usepackage{float}
\usepackage{esint}

\usepackage{color}
\usepackage[english]{babel}
\usepackage[T1]{fontenc}
\usepackage[utf8]{inputenc}
\usepackage[font=small,labelfont=bf]{caption}
\usepackage{mwe}    
\usepackage{subcaption}

\def\prl{{Phys. Rev. Lett.}}

\def\pre{{Phys. Rev. E}}

\def\pop{Phys. Plasmas}
\def\pof{Phys. Fluids}
\def\pofb{Phys. Fluids B}
\def\jfm{J. Fluid Mech.}
\def\pr{Phys. Rev.}
\def\jetp{Sov. Phys. JETP Lett.}
\def\jcp{J. Comput. Phys.}

\begin{document}

\title{Deciphering the Kinetic Structure of Multi-Ion Plasma Shocks}

\author{Brett D. Keenan}
\email{keenan@lanl.gov}
\author{Andrei N. Simakov}
\author{Luis Chac\'{o}n}
\author{William T. Taitano}
\affiliation{Los Alamos National Laboratory, Los Alamos, New Mexico 87545, USA}


\begin{abstract}
Strong collisional shocks in multi-ion plasmas are featured in many high-energy-density environments, including Inertial Confinement Fusion (ICF) implosions. However, their basic structure and its dependence on key parameters (e.g., the Mach number and the plasma ion composition) are poorly understood, and controversies in that regard remain in the literature. Using a high-fidelity Vlasov-Fokker-Planck code, iFP, and direct comparisons to multi-ion hydrodynamic simulations and semi-analytic predictions, we critically examine steady-state planar shocks in D-$^3$He plasmas and put forward a resolution to these controversies.
\end{abstract}

\maketitle

{\it Introduction}.---Strong shocks in multi-ion plasmas are key to a number of high-energy density settings. One prominent example is laser driven Inertial Confinement Fusion (ICF) implosions, which rely upon strong shocks for initial compression and heating of the fuel. Consequently, multi-ion effects (e.g., mass diffusion and temperature separation) and kinetic effects associated with shocks may crucially affect the performance of ICF implosions \citep{rosenberg14, rosenberg14a, rinderknecht14, rosenberg15, rinderknecht15}.  

ICF implosions are normally simulated with radiation-hydrodynamics (rad-hydro). Such a treatment is valid for $N_K \ll 1$, where $N_K$ (the Knudsen number) is the ratio of the constituent ions' mean-free-path to a characteristic gradient length scale. From simple hydro estimates, as the Mach number, $M$, of a collisional shock increases, so should $N_K$ \citep{zeldovich67}. Consequently, the hydrodynamic treatment is formally valid only for weak shocks with $M - 1 \ll 1$ \citep{jukes57, mott-smith51}, where $N_K \sim 2(M-1)$. Thus, a kinetic treatment is required for moderate ($M-1 \sim 1$) and strong ($M \gg 1$) shocks. Despite this limitation, the structure of steady-state planar shocks in single species plasmas has been studied in the hydro limit by multiple authors \citep{jukes57, shafranov57, jaffrin64, grewal73}.  

Initial kinetic studies of strong shocks in a single-species plasma employed the Mott-Smith ansatz. The Mott-Smith approach \citep{greywall75} admitted a solution to the Vlasov-Fokker-Planck (VFP) equations, by assuming that the particle velocity space distribution in a strong shock has a bi-Maxwellian form determined by the upstream and downstream conditions. These studies concluded that the kinetic shock width is considerably greater than the hydro equivalent \citep{mott-smith51, greywall75, abe75}, and is expected to grow with $M$ \citep{tidman58, muckenfuss60, greywall75}. Using a finite electron thermal conductivity, Ref.\ \citep{greywall75} predicted finite asymptotic growth and saturation of the ion shock width (normalized to a downstream mean-free-path) as $M \rightarrow \infty$. The physical explanation is that collisionality is effectively reduced at sites of short gradient scales (compared to the ion mean-free-path) in the kinetic regime. Thus, hot ions moving from the downstream toward the upstream extend the shock width beyond its hydro limits. This broadening of the shock width for $M \gg 1$ was later observed in FPION VFP simulations of a hydrogen plasma shock \citep{casanova91, vidal93}.

In contrast to the Mott-Smith prediction in Ref.\  \citep{greywall75}, researchers using the hybrid Particle-in-Cell (PIC) code LSP found that the shock width (normalized to the ion-ion mean-free-path in the downstream) reaches a maximum at $M \sim 6$, and monotonically decreases for larger $M$ \citep{bellei14}. 

It is worth noting that these studies typically considered single-ion plasmas. Hence, the structure of a collisional shock in a multi-ion plasma is a largely unexplored problem, and some controversies exist in the sparse multi-ion literature so far \citep{larroche12, bellei14, bellei14a}. FPION simulations for multi-ion (planar) plasma shocks in an equimolar mixture of deuterium and helium-3 \citep{larroche12} predicted deuterium temperature profiles overcoming electron temperature ones in the entirety of the electron pre-heat layer. In contrast, earlier single ion-species FPION studies \citep{vidal93} and multi-fluid analyses \citep{glazyrin16} found an ion temperature lower than the electron temperature in the pre-heat layer. In principal, one would expect that the electron temperature will precede any other species; given that the electrons have a far greater thermal velocity than the ions, and because the principal heating mechanism for the ions in the pre-heat layer is thermal exchange with the hotter electrons \citep{zeldovich67}.

In this Letter, we resolve these controversies by systematically studying shocks in D-$^3$He plasma with both a VFP and a multi-ion hydro approaches \citep{simakov17}. For this study, we employ a state-of-the-art VFP code, iFP. This code is fully mass, energy, and momentum conserving; it is also adaptive and well verified \citep{taitano15, taitano16, yin16, taitano17, taitano17a}. iFP treats ions fully kinetically, resolving both species within their own separate velocity-spaces, while simultaneously solving the quasi-neutral fluid equations for electrons \citep{simakov14}. 

Our multi-ion hydro code is grounded in a multi-species generalization of the Braginskii equations \citep{simakov14, simakov16, simakov16a, simakov17}. While it includes full ion diffusion, it assumes that both ion species have the same temperature. This is a valid assumption for weak to intermediate strength shocks with $M - 1 \simeq 1$, since temperature separation is a higher-order effect in $N_K \ll 1$ (for details, see Ref. \citep{simakov17}), but not for $M \gg 1$. The hydro code has been benchmarked against analytical shock profiles for $M -1 \ll 1$ \citep{simakov17}.

As we will show, the shock width we observe is consistent with early Mott-Smith studies. Furthermore, it also depends upon the relative concentration of the lighter plasma species, $c \equiv \rho_l/\rho$, where $\rho_l$ is the mass density of the lighter species, and $\rho$ is the total mass density of the plasma.

\begin{figure}
\begin{subfigure}[b]{.45\textwidth}
  \centering
  \includegraphics[width=1\linewidth]{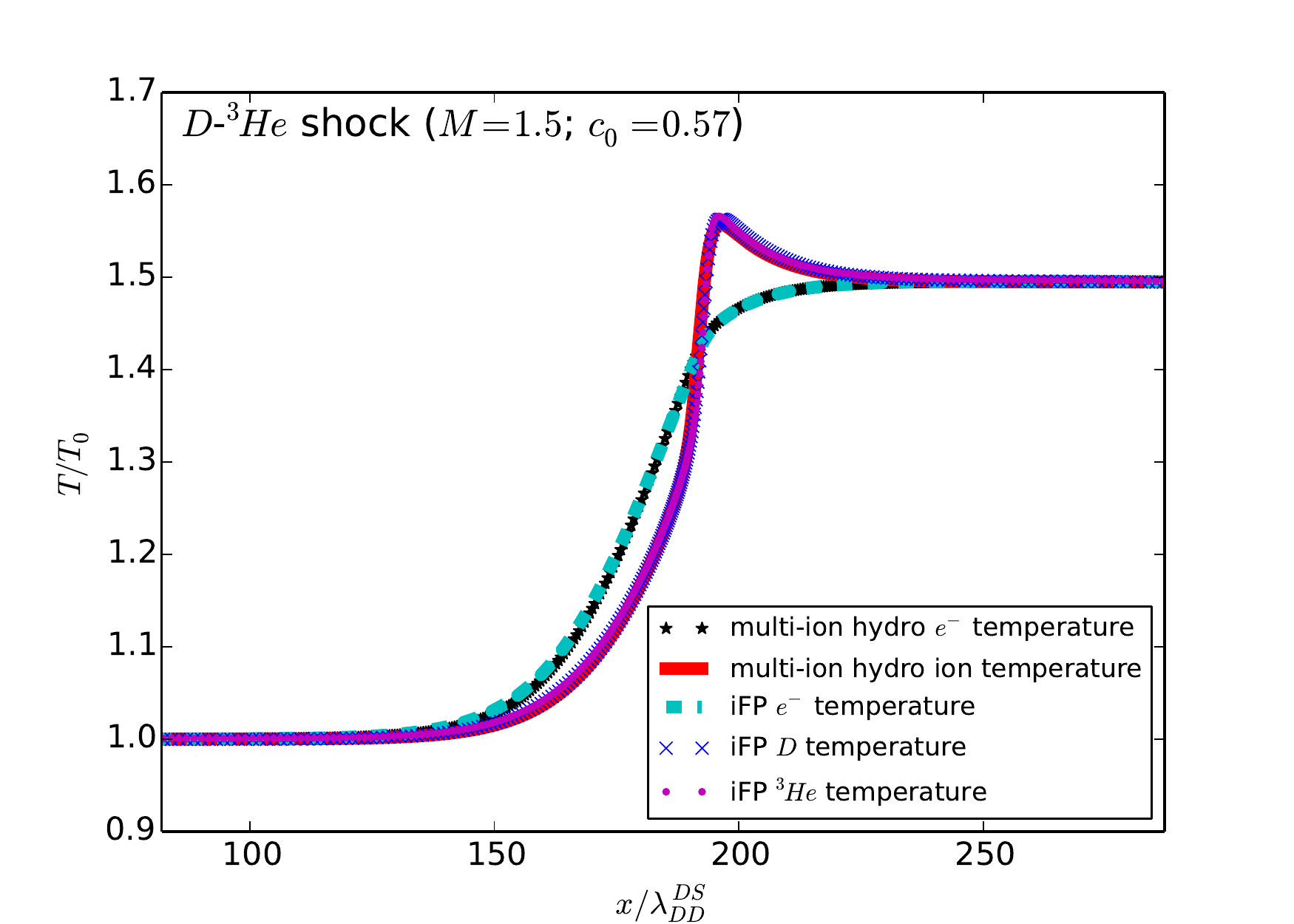}
  \caption{}
  \label{mach1.5_temp}
\end{subfigure}%
\newline
\noindent
\begin{subfigure}[b]{.45\textwidth}
  \centering
  \includegraphics[width=1\linewidth]{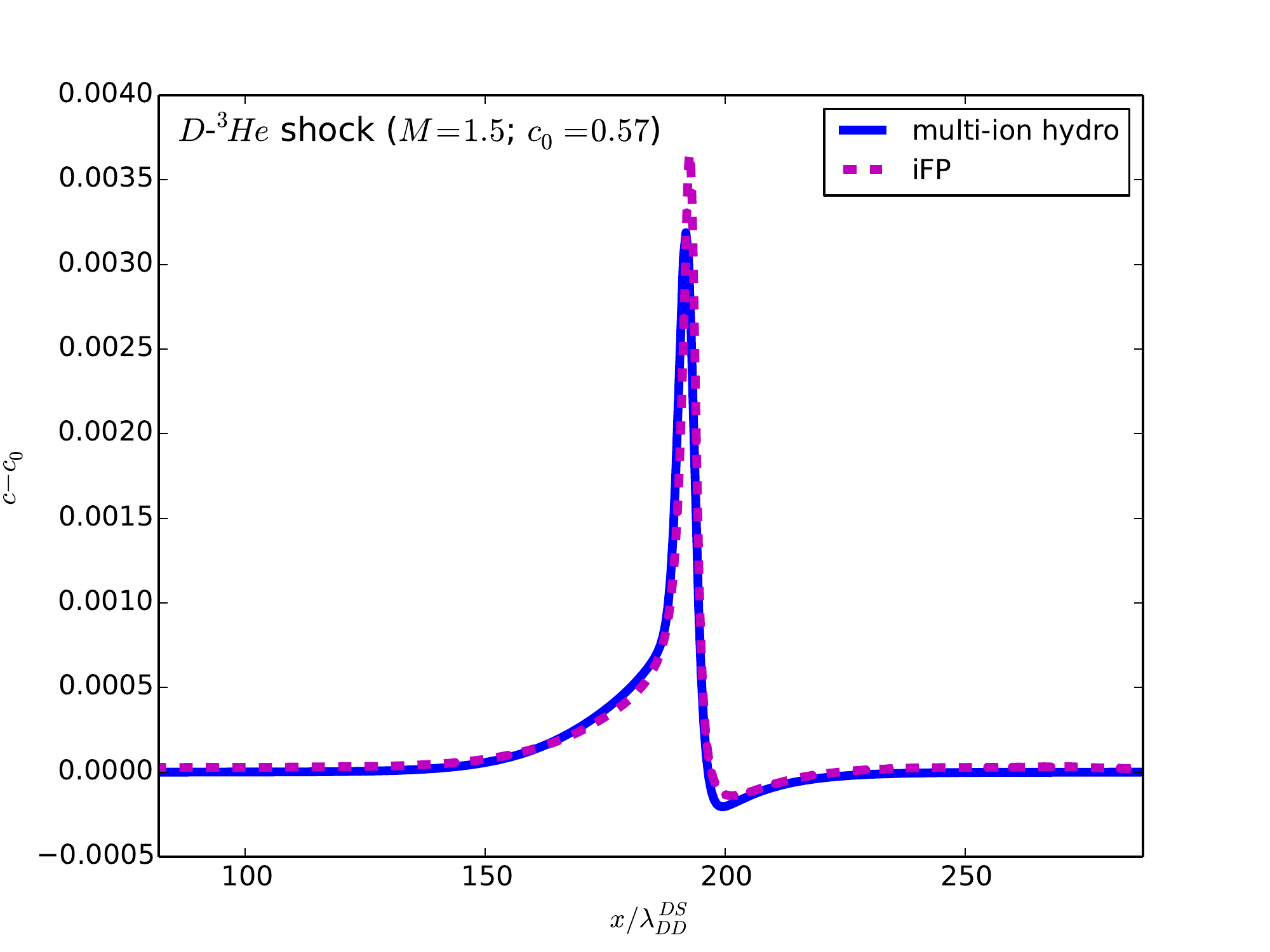}
  \caption{}
  \label{mach15_c-c0}
\end{subfigure}
\caption{(Color online). Electron and ion temperature profiles (a) and the change in deuterium concentration, $c-c_0$, (b) for an $M = 1.5$ shock.}
\end{figure}

{\it Intermediate Strength Shocks}.---Weak shocks are accurately described with multi-ion hydrodynamics \citep{simakov17}. We begin by demonstrating that iFP produces correct results in this limit. Recovering the hydro limit is difficult for Fokker-Planck codes, and therefore, this is a challenging test of iFP's capabilities. To this end, we compare iFP results to our multi-ion hydrodynamic simulations. In Fig.\ \ref{mach1.5_temp}, we show electron and ion temperature comparisons. The initial mass concentration of deuterium (in the upstream, i.e., the left side of the figure) is $c_0 \equiv m_Dn_D/\rho = 0.57$, where $m_D$ and $n_D$ are the deuterium mass and number densities, respectively. The x-axis is normalized to the DD mean-free-path in the downstream, $\lambda_{DD}^{DS}$, and $T_0$ is the upstream temperature. Although an $M = 1.5$ shock is not strictly ``weak'', iFP and our multi-component hydro code demonstrate superb agreement. Note that, herein, all iFP and hydrodynamic simulations assume a constant Coulomb logarithm of 10 for all species.

Figure \ref{mach1.5_temp} also displays a fundamental feature of intermediate and strong plasma shocks: the electron pre-heat layer. This is the region immediately to the right of the upstream, where the electron temperature sharply changes ahead of the ion temperatures, owing to the much smaller electron mass and greater thermal conductivity.
\newline
\indent
In Fig.\ \ref{mach15_c-c0}, we confirm very good agreement in D concentration change across the shock front between iFP and the multi-ion hydro simulation prediction for $M = 1.5$. Since the change in the deuterium mass concentration ($c-c_0$, where $c_0$ is the upstream concentration) is more sensitive to $N_K$ than the temperatures \citep{simakov17}, larger differences between the hydro and kinetic results can be better appreciated.

\begin{figure}
\begin{subfigure}[b]{.45\textwidth}
  \centering
  \includegraphics[width=1\linewidth]{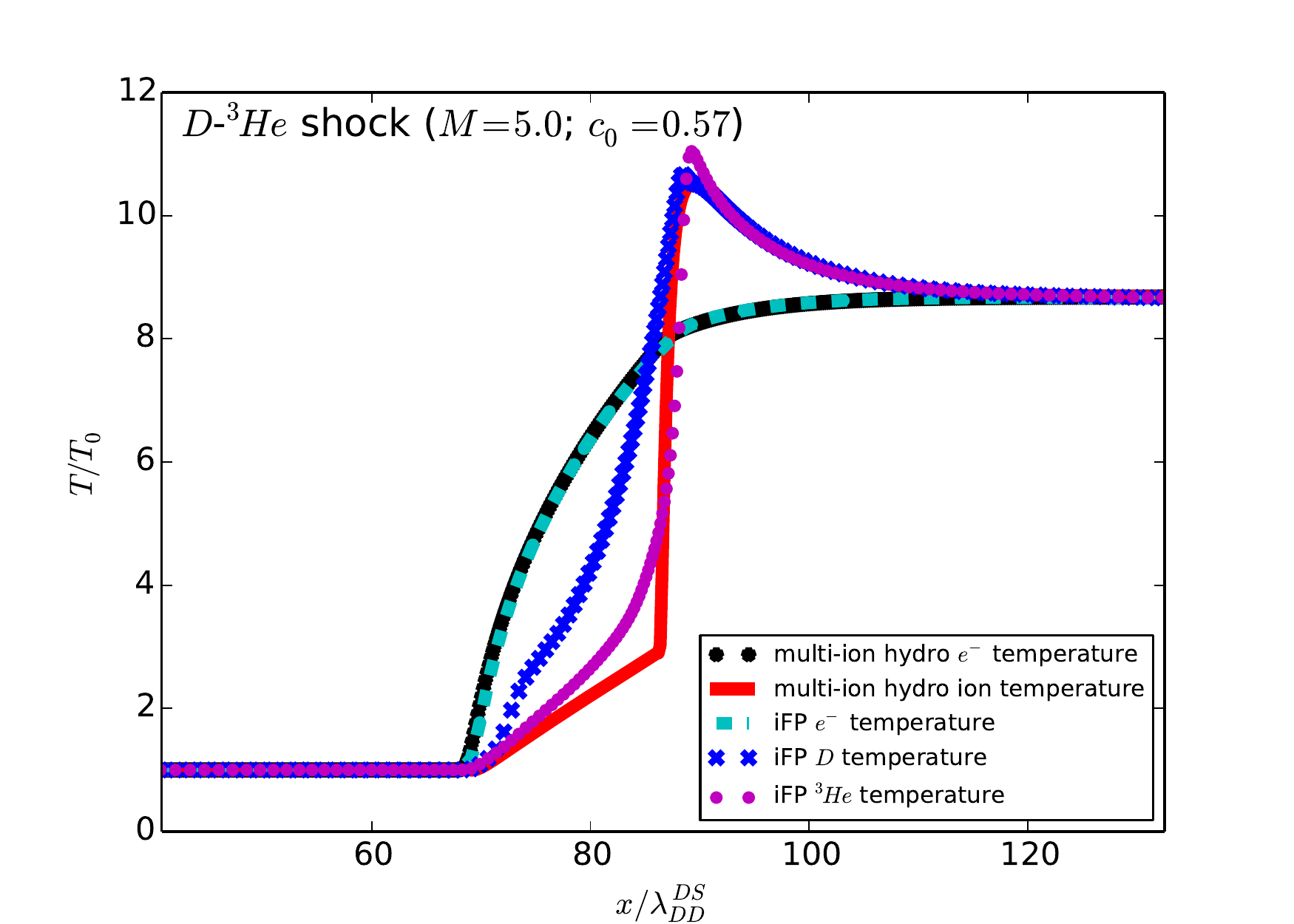}
  \caption{}
  \label{mach5_temp}
\end{subfigure}%
\newline
\noindent
\begin{subfigure}[b]{.45\textwidth}
  \centering
  \includegraphics[width=1\linewidth]{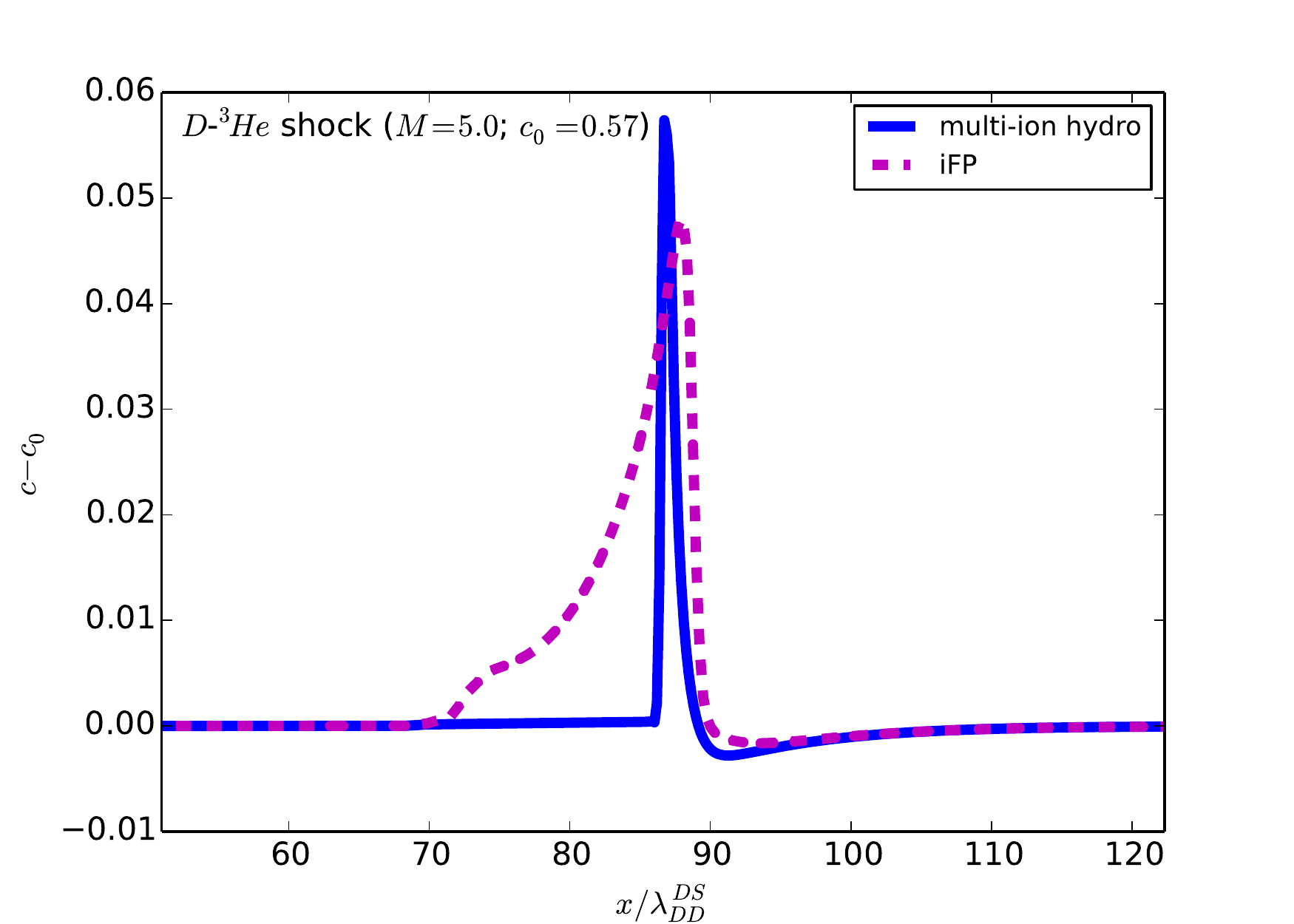}
  \caption{}
  \label{mach5_c-c0}
\end{subfigure}
\caption{(Color online). Temperature profiles (a) and deuterium enrichment (b) for an $M = 5$ shock.}
\end{figure}

\begin{figure}
\begin{subfigure}[b]{.45\textwidth}
  \centering
  \includegraphics[width=1\linewidth]{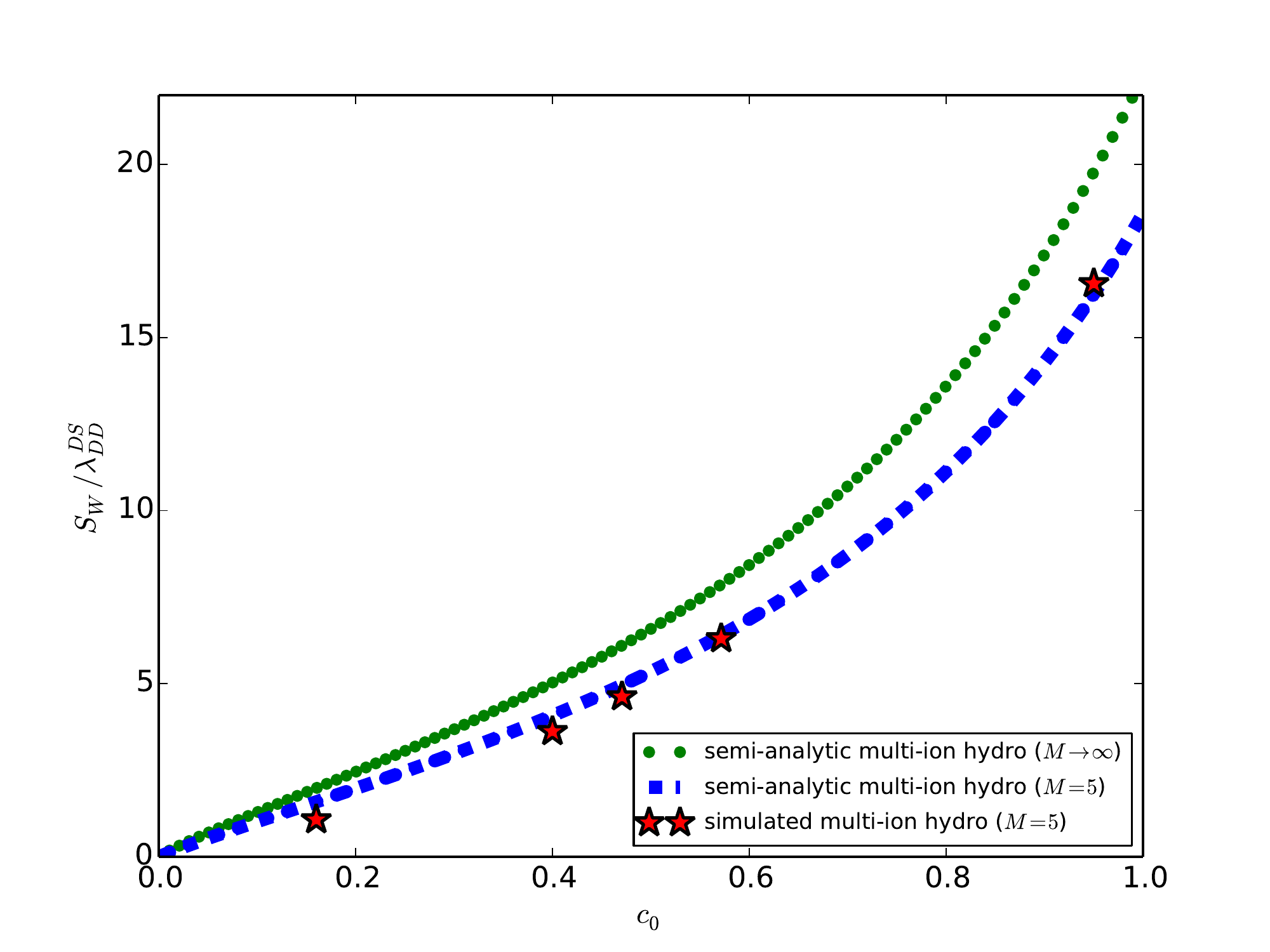}
  \caption{}
  \label{sw_hydro}
\end{subfigure}%
\newline
\noindent
\begin{subfigure}[b]{.45\textwidth}
  \centering
  \includegraphics[width=1\linewidth]{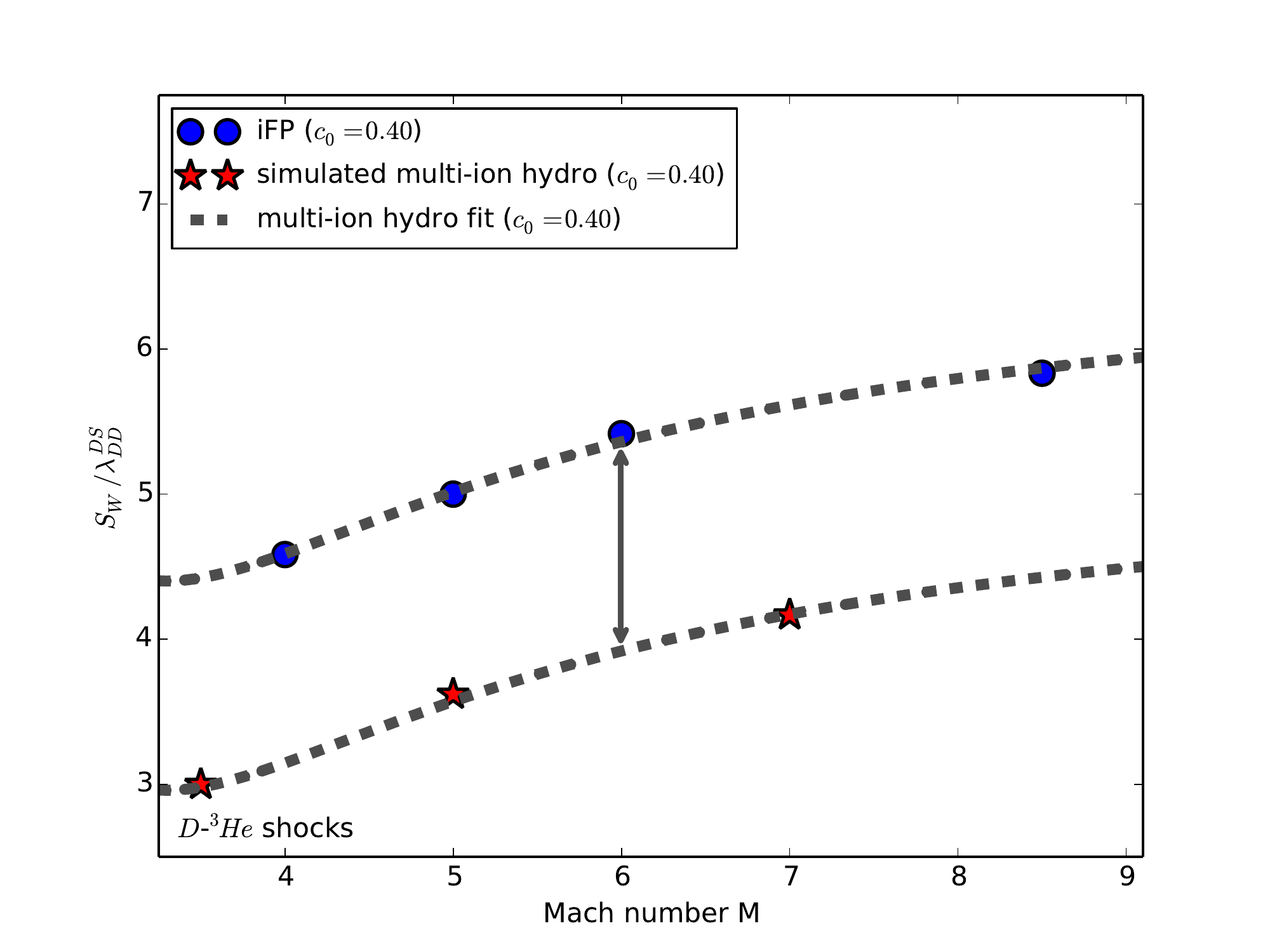}
  \caption{}
  \label{sw_ifp}
\end{subfigure}
\caption{Semi-analytic multi-ion hydro shock width (for $M \gg 1$ and $M = 5$) vs.\ $c_0$ (a), and the ion shock width vs.\ Mach number for $c_0 = 0.40$ (b). Included in (a) are results from full multi-ion hydro simulations (for $M = 5$).}
\end{figure}

\begin{figure}
\includegraphics[angle = 0, width = 1\columnwidth]{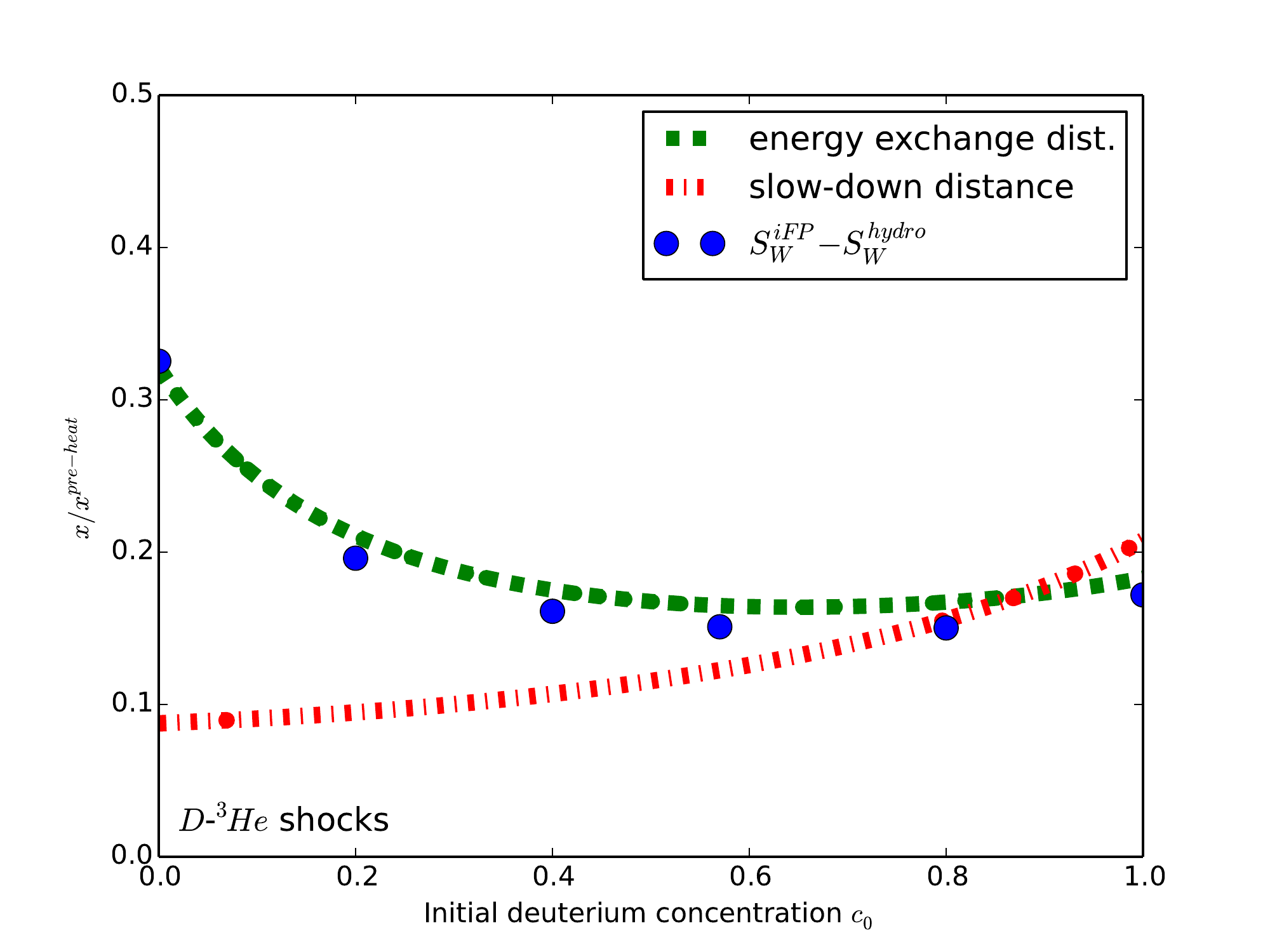}
\vskip-0.1cm
\caption{(Color online). Simulation results for $S_W^{iFP} - S_W^{hydro}$ normalized to the length of electron pre-heat layer vs.\ $c_0$ and two proposed theoretical models.}
\label{c0_dist}
\end{figure}

{\it Strong Shocks}.---Next, we consider strong shocks. The structure of a strong (hydro) plasma shock is well known \citep{jukes57, shafranov57, jaffrin64, grewal73}. There are three principal regions: 1) an electron pre-heat ``pedestal'' layer where the electron temperature exceeds the ion temperature, 2) the imbedded/compression ion shock, and 3) an equilibration layer where the electron and ion temperatures relax to the downstream values. The widths of regions 1) and 3) are $\sim \sqrt{\frac{m_i}{m_e}}\lambda_{ii}^{DS}$ (where $m_i$ and $m_e$ are the ion and electron masses, respectively, and $\lambda_{ii}^{DS}$ is the downstream ion-ion mean-free-path), whereas the width of 2) is a few ion-ion mean-free-paths. With regard to the length of the pre-heat layer, we have found from hydro simulations that a more accurate estimate is given by $x^{pre-heat} \sim \lambda_{ee}\frac{v_{the}}{u_0}$, where $\lambda_{ee}$ is the downstream electron-electron mean-free-path, $v_{the}$ is the electron thermal velocity in the downstream, and $u_0$ is the shock velocity. It is easy to show that $\lambda_{ee}\frac{v_{the}}{u_0} \sim \sqrt{\frac{m_i}{m_e}}\lambda_{ii}^{DS}$.
\newline
\indent
In the following plots, we feature an $M = 5$ shock, since our simulations indicate that the essential structure of a plasma kinetic shock is adequately captured here; i.e., our results for higher $M$ are qualitatively similar.
\newline
\indent
In Fig.\ \ref{mach5_temp}, we show for an $M = 5$ D-$^3$He shock a considerable kinetic enhancement of the ion temperature in the pre-heat layer and at the shock front vs.\ the hydro simulations. According to Fig.\ \ref{mach5_c-c0}, the deuterons penetrate a greater depth into the upstream due to their higher thermal velocity with respect to $^3$He. In contrast, the corresponding multi-component hydro result shows a sharp cutoff of the ion enrichment at the shock front, corresponding to the sharp gradient in the ion temperature at that point. Consequently, the build up of deuterium in the upstream is a purely kinetic effect. Note that the multi-ion hydro prediction for the ion temperature shows the pedestal feature, which ends abruptly at the sharp cutoff point for $c-c_0$. The pedestal structure, which also appears in the density profiles (not shown), is smoothed in kinetic shocks. Moreover, the D temperature is higher than the $^3$He temperature in the pre-heat layer.
\newline
\indent
{\it Shock Width}. --- As mentioned previously, various claims have been made about the width of the shock for $M \gg 1$. Reference \citep{bellei14} defined the shock width ($S_W$) as the length over which the ion density increases from 1.2 times its upstream value, $\rho_0$, to 0.9 times its downstream value, $\rho_1$ (normalized to the ion-ion mean-free-path in the downstream). This definition, which is not unique, is attractive for strong shocks, not only because it is largely insensitive to electron heat flux limiters \citep{bellei14}, but also because this width falls short of the density pedestal ahead of the shock front (a purely hydro-like feature). For a meaningful comparison, we adopt the same definition. 
\newline
\indent
From Ref.\ \citep{simakov17}, a hydrodynamic two-ion, stationary, plasma shock may be described by two equations. Firstly, we have the electron energy equation:
\begin{equation}     
\partial_{\hat{x}}\left(\frac{3}{2}\hat{p}_{e0}\hat{T}_e - \hat{\kappa}_e\partial_{\hat{x}}\hat{T}_e\right) + \hat{p}_{e0}\frac{\hat{T}_e}{\hat{V}}\partial_{\hat{x}}\hat{V} = \hat{\nu}_{ei}\left(\hat{T}_i - \hat{T}_e\right),
\label{dTedx}
\end{equation}
where $\hat{V} \equiv \rho_0/\rho$, $\hat{p}_{e0}$ is the electron pressure in the upstream normalized by the total upstream pressure, $\hat{\kappa}_e$ is the normalized electron thermal conductivity coefficient, $\hat{T}_i$ is the single ion temperature, $\hat{\nu}_{ei}$ is an electron-ion energy exchange frequency, $\hat{x}$ is the distance normalized to the DD mean-free-path in the upstream, and all other quantities are normalized to their respective upstream values. Next, we have an equation for the ion mass density:
\begin{equation}     
2\gamma M^2\left(\hat{V} - 1\right)\left(\hat{V}_1 - \hat{V}\right) + \frac{3}{2}\hat{\eta}\hat{V}\partial_{\hat{x}}\hat{V} \approx \hat{\kappa}_e\partial_{\hat{x}}\hat{T}_e,
\label{dVdx0}
\end{equation}
where $\hat{\eta}$ is the normalized ion viscosity cofficient, $\hat{V}_1 \equiv \rho_0/\rho_1$, $\gamma = 5/3$ is the adiabatic index. Formally, the right-hand-side of Eq.\ (\ref{dVdx0}) includes the ion heat and differential mass diffusion fluxes, but our hydro simulations indicate that the electron thermal conductivity term, $\hat{\kappa}_e\partial_{\hat{x}}\hat{T}_e$, dominates. 
\newline
\indent
The electron and ion temperatures inside the shock front scale as $M^2$, since they are of order the downstream temperature. Next, we note that $\hat{\nu}_{ei}(\hat{T}_i - \hat{T}_e)$ in Eq.\ (\ref{dTedx}) scales as $1/M^2$, since $\hat{\nu}_{ei} \propto M^{-4}$. This expression is generally smaller than the left-hand-side of the equation, owing to the fact that $\hat{\nu}_{ei}$ contains a factor of $\sqrt{\frac{m_e}{m_D}}$ and the energy exchange between ions and electrons is not the primary heating mechanism within the imbedded shock. For this reason, we ignore the energy exchange term in Eq.\ (\ref{dTedx}), to obtain:
\begin{equation}     
\left[\frac{3}{2} + \text{ln}(\hat{V})\right]\hat{p}_{e0}\hat{T}_{e-}  - \hat{\kappa}_e\partial_{\hat{x}}\hat{T}_e \approx {\it const},
\label{dTedx_approx}
\end{equation}
where we have used the fact that the electron temperature within the shock, denoted by $\hat{T}_{e-}$, is approximately constant \citep{shafranov57}. The integration constant is effectively zero, as follows from the upstream boundary condition.  Additionally,  $1/4 \leq \hat{V} \leq 1$.  Given these considerations, and the fact that $\hat{T}_{e-} \propto M^2$, we conclude that $\hat{\kappa}_e\partial_{\hat{x}}\hat{T}_e$  must also scale as $M^2$.
\newline
\indent
We now turn our attention to Eq.\ (\ref{dVdx0}). The coefficient of ion viscosity, $\hat{\eta}$, scales as $M^6$. We may re-write Eq.\ (\ref{dVdx0}) as:
\begin{equation}     
\frac{d\hat{x}}{d\hat{V}} \approx \frac{\frac{3}{2}\hat{\eta}\hat{V}}{\hat{\kappa}_e\partial_{\hat{x}}\hat{T}_e - 2 \gamma M^2(\hat{V}-1)(\hat{V}_1-\hat{V})}, 
\label{Sw_def_integrand}
\end{equation}
from which we may conclude that $d\hat{x}/d\hat{V}$ scales as $M^4$, which is the Mach number dependence found in Refs.\ \citep{tidman58, abe75} for strong shocks using the Mott-Smith ansatz. Normalizing this to the downstream mean-free-path introduces a factor of $1/M^4$, indicating that the normalized shock width:
\begin{equation}     
S_W \equiv \frac{1}{\hat{V}_1{\hat{T}_1}^2}\int_{\hat{V} = \frac{1}{1.2}}^{\hat{V} = \frac{\hat{V}_1}{0.9}}\frac{d\hat{x}}{d\hat{V}}d\hat{V},
\label{Sw_def}
\end{equation}
does not scale with $M$, and therefore reaches a finite asymptotic value as $M \rightarrow \infty$, which is in agreement with Ref.\  \citep{greywall75}. 
\newline
\indent
To integrate this equation, we first note that the electron temperature within the imbedded shock is approximately constant. For the portion of the pre-heat layer nearest to the upstream, $\text{ln}(\hat{V}) \approx \hat{\nu}_{ei}(\hat{T}_i - \hat{T}_e) \approx 0$, and thus we may directly obtain the electron temperature in the pre-heat layer from Eq.\ (\ref{dTedx}) as \citep{zeldovich67}:
\begin{equation}     
\hat{T}_e(\hat{x}) \approx \left[\frac{15}{4}\frac{\hat{p}_{e0}}{\hat{\kappa}_{e0}}\left(\hat{x}-\hat{x}_0\right) + 1\right]^{\frac{2}{5}},
\label{Te_preheat}
\end{equation}
where $\hat{\kappa}_{e0} \equiv \hat{\kappa}_e |_{\hat{x} = \hat{x}_0}$, and $\hat{x}_0$ is the position of the upstream edge of the pre-heat layer. To obtain $\hat{T}_{e-}$, we evaluate Eq.\ (\ref{Te_preheat}) at the location of the imbedded shock, which is at $\hat{x} - \hat{x}_0 = \hat{x}^{pre-heat} \sim \lambda_{ee}v_{the}/(u_0\lambda^{US}_{DD})$, where $\lambda^{US}_{DD}$ is the ion-ion mean-free-path in the upstream. The exact value of $\hat{x}^{pre-heat}$, which depends on $c_0$, $M$, etc., is unknown. Consequently, we slightly tweak $\hat{T}_{e-}$ to best fit the results from full multi-ion hydro simulations. An expression for $\hat{T}_{e-}$ allows us to estimate \citep{simakov17} $\hat{\eta} \propto \hat{T}_i^{5/2}$ and $\hat{\kappa}_e\partial_{\hat{x}}\hat{T}_e$, and we are then able to obtain the hydro shock width as a function of $M$ and $c_0$ using Eqs.\ (\ref{Sw_def_integrand}) and (\ref{Sw_def}).
\newline
\indent
In Fig.\ \ref{sw_hydro}, we present this result for $M \rightarrow \infty$ and $M = 5$ as functions of $c_0$. The semi-analytic curve for $M = 5$, the blue dashed line, matches the multi-ion hydro simulation points, shown in red stars, very closely. 
\newline
\indent
Figure \ref{sw_ifp} shows the shock width as a function of Mach number from full multi-ion hydro simulations as red stars for $c_0 = 0.40$, along with a gray dashed fit curve. The figure also shows the corresponding iFP shock width. Overlapping the latter points is the fit for the hydro regime, translated upward by a fixed amount that depends on $c_0$, but not $M$. We see that the kinetic shock width, for $M \gg 1$, is simply the multi-ion hydro shock width plus a correction. It follows that the kinetic shock width asymptotes to a constant as $M \rightarrow \infty$. Thus, Fig.\ \ref{sw_ifp} is consistent with the Mott-Smith results from Refs. \citep{tidman58, abe75, greywall75}, but is at odds with Ref.\ \citep{bellei14}, which predicted that the shock width decreases for $M \gtrsim 6$. 
\newline
\indent
The kinetic extension of the shock width is simply due to hot downstream ions penetrating upstream beyond the ion density pedestal \citep{vidal93}. We may quantify this effect by plotting $S_W^{iFP} - S_W^{hydro}$ vs.\ $c_0$, as in Fig.\ \ref{c0_dist}, normalized to the length of the electron pre-heat layer. Here, the dots are from simulations, and the green dashed line is a characteristic ion energy relaxation distance. 

The ion energy relaxation distance is obtained by considering a flow of downstream deuterons into a colder pedestal plasma in the upstream. Kinetic effects are most prominent in the pedestal, since this is the site of the sharpest gradients in temperature, density, etc. Additionally, this definition of the distance is independent of Mach number for $M \gg 1$ \citep{grewal73}. 
\newline
\indent
This distance can be estimated by considering a characteristic time for hot particles, $\alpha$, to exchange energy with colder background plasma particles, $\beta$ \citep{trubnikov}:
\begin{equation}     
\frac{1}{\tau^\alpha_\epsilon} \equiv \sum_\beta \frac{1}{\tau^{\alpha/\beta}_\epsilon},
\label{tau_def}
\end{equation}
where:
\begin{equation}     
\tau_\epsilon^{\alpha/\beta} = \frac{\tau_1^{\alpha/\beta}}{4\mu(x_\beta)/x_\beta},
\label{tau_eps_def}
\end{equation}
$x_\beta = \left(\frac{m_\beta}{m_\alpha}\right)\frac{T_\alpha}{T_\beta}$, $\mu(x) = \frac{2}{\sqrt{\pi}}\int_0^x \sqrt{t}e^{-t}dt$, $\tau_1^{\alpha/\beta} = \frac{\sqrt{m_\alpha}}{\pi\sqrt{2}e_\alpha^2e_\beta^2}\frac{T_\alpha^{3/2}}{n_\beta\text{ln}(\Lambda)}$, with $\text{ln}(\Lambda)$ the Coulomb logarithm, $T_\alpha$ the downstream $D$ temperature, $n_\beta$ and $T_\beta$ the number densities and temperatures of the plasma species in the pedestal region, respectively, and $m_\alpha$, $m_\beta$, $e_\alpha$, and $e_\beta$ the corresponding masses and charges. Finally, we define the ion energy relaxation distance for D as:
\begin{equation}     
d_\epsilon \equiv u_0\tau_\epsilon^{D}.
\label{dist_def}
\end{equation}
\newline
\indent
Reference \citep{vidal93} proposed that the kinetic extension of a plasma shock roughly corresponds to a characteristic slow-down distance, which is defined similarly to that in Eq.\ (\ref{dist_def}) but in terms of the slow-down time \citep{trubnikov}:
\begin{equation}     
\tau_s^{\alpha/\beta} = \frac{\tau_1^{\alpha/\beta}}{\left[1 + \frac{m_\alpha}{m_\beta}\right]\mu(x_\beta)}.
\label{slow_def}
\end{equation}
Using the same definitions of all plasma quantities, we have plotted the slow-down distance in Fig.\ \ref{c0_dist} as the red dash-dotted line. Evidently, the extension of the kinetic shock width is more accurately represented by the total ion energy relaxation distance rather than the slow-down distance. Moreover, from Fig.\ \ref{c0_dist} it is clear that the ion energy exchange distance is less than the electron pre-heat layer width for all $c_0$,  suggesting that the ion temperature cannot overcome the electron temperature in the upstream edge of the pre-heat layer.
\newline
\indent
{\it Discussion}.---We have performed high-fidelity shock simulations with the VFP code iFP. In the weak shock regime (hardest for VFP codes), the iFP results show excellent agreement with multi-ion hydro simulations, underscoring iFP's accuracy and reliability. 

In the strong shock regime, we found that kinetic effects saturate for $M \gg 1$ (i.e., the kinetic extension of the shock width, normalized to a downstream mean-free-path, becomes independent of  the Mach number). Moreover, the asymptotic value strongly depends on the upstream deuterium concentration. This is true of both the multi-ion hydro and kinetic shock widths, with the latter exceeding the former by a characteristic ion energy exchange distance. Our findings, while consistent with early Mott-Smith predictions \citep{greywall75} and Fokker-Planck results \citep{vidal93}, conflict with the LSP (PIC) predictions from Ref. \citep{bellei14}.

Additionally, we have found that the ion temperature never exceeds the electron temperature in the pre-heat layer, contradicting a recent study with FPION \citep{larroche12}, but in agreement with an earlier FPION study \citep{vidal93}, and a study using a multi-fluid model \citep{glazyrin16}. Rather, the kinetic multi-ion shock width is extended beyond the hydro equivalent by a characteristic energy exchange distance, which is less than the length of the electron pre-heat layer.

\begin{acknowledgments}

{\it Acknowledgments}.--This work was supported by the Los Alamos National Laboratory LDRD Program, Metropolis Postdoctoral Fellowship for W.T.T., and used resources provided by the Los Alamos National Laboratory Institutional Computing Program. Work performed under the auspices of the U.S.\ Department of Energy National Nuclear Security Administration under Contract No.\ DE-AC52-06NA25396.

\end{acknowledgments}

\end{document}